\documentclass[12pt,letterpaper]{article}

\usepackage{epsfig}
\usepackage{pstricks}
\usepackage{fancyhdr}
\usepackage{graphicx}
\usepackage{multind}
\usepackage[square,comma,numbers]{natbib}
\usepackage{dcolumn}
\usepackage{bm}
\usepackage{mathrsfs} 
\usepackage{xspace}
\usepackage{array,amsfonts,amssymb,amsmath,longtable}

\newcommand{\sss}{\scriptscriptstyle}

\begin{document}
\title{
Hadronic Higgs Production with Heavy Quarks at the Tevatron and 
the LHC
\footnote{Prepared for 
  the TeV4LHC Higgs Working Group Report.}}
\author{S.~Dawson\footnote{dawson@bnl.gov} and 
C.~B.~Jackson\footnote{jackson@quark.phy.bnl.gov}\\
{\em\small Physics Department, Brookhaven National  Laboratory, Upton,N.Y.~~ USA}\\
L.~Reina\footnote{reina@hep.fsu.edu}\\
{\em
\small 
Physics Department, Florida State University,
Tallahassee, FL 32306-4350, USA}\\
D.~Wackeroth
\footnote{dow@ubpheno.physics.buffalo.edu}\\
{\em
\small Department of Physics, SUNY at Buffalo,
Buffalo, NY 14260-1500, USA}}

\date{}
\maketitle
\begin{abstract}
We review the status of the QCD corrected cross sections and kinematic distributions
for the production of a Higgs boson in association with top quark or bottom
quark pairs at the Fermilab Tevatron and at the LHC.  Results for $b {\overline b}
H$ production are presented in the Minimal Supersymmetric Model, where the
rates can be greatly enhanced relative to the Standard Model rates. We place particular 
emphasis on theoretical uncertainties due to renormalization and 
factorization scale dependence and on the uncertainties coming from the Parton Distribution Functions. 
\end{abstract}


%
\label{sec:djrw}
\section{Introduction}
\label{sec:djrw_intro}
A light Higgs boson is preferred by precision fits of the Standard
Model (SM) and also theoretically required by the Minimal
Supersymmetric extension of the Standard Model (MSSM).  The production
of a Higgs boson in association with a heavy quark and antiquark pair,
both $t\bar{t}$ and $b\bar{b}$, at the Tevatron and the Large Hadron
Collider (LHC) will be sensitive to the Higgs-fermion couplings and
can help discriminate between models.

The associated production of a Higgs boson with a pair of $t\bar{t}$
quarks has a distinctive signature and can give a direct measurement
of the top quark Yukawa coupling.  This process is probably not
observable at the Tevatron, but will be a discovery channel at the LHC
for $M_h < 130$~GeV.  The associated production of a Higgs boson with
a pair of $b\bar{b}$ quarks has a small cross section in the Standard
Model, and can be used to test the hypothesis of enhanced bottom quark
Yukawa couplings in the MSSM with large values of $\tan\beta$. Both
the Tevatron and the LHC will be able to search for enhanced
$b\bar{b}h$ production, looking for a final state containing no bottom
quarks (inclusive production), one bottom quark (semi-inclusive
production) or two bottom quarks (exclusive production).

The rates for $t {\overline t} h$ production at the Tevatron and the
LHC have been calculated at NLO QCD several years
ago\cite{Reina:2001sf,Beenakker:2001rj,Reina:2001bc,Beenakker:2002nc,
Dawson:2002tg,Dawson:2003zu}.  The theoretical predictions for
$b\bar{b}h$ production at hadron colliders involve several subtle
issues, and depend on the number of bottom quarks identified in the
final state. In the case of no or only one tagged bottom quark there
are two approaches available for calculating the cross sections for
$b\bar{b}h$ production, called the four flavor number schemes
(4FNS)\cite{Dittmaier:2003ej,Dawson:2004sh} and five flavor number
scheme (5FNS)\cite{Maltoni:2003pn}.  The main difference between these
two approaches is that the 4FNS is a fixed-order calculation of QCD
corrections to the $gg$ and $q\bar{q}$-induced $b\bar{b}h$ production
processes, while in the 5FNS the leading processes arise from $bg$
($\bar{b}g$) and $b\bar{b}$ initial states and large collinear
logarithms are resummed using a pertubatively defined bottom quark
Parton Distribution Function (PDF). Very good agreement is found for
the NLO QCD corrected cross sections for $b\bar b$ Higgs associated
production when the two schemes are
compared\cite{Dawson:2005vi,Campbell:2004pu}.

In the following sections, we present numerical results at NLO QCD for
$t {\overline t} h$ and $b {\overline b} h$ production at the Tevatron
and the LHC.  If not stated otherwise, numerical results have been
obtained in the 4FNS. We emphasize theoretical uncertainties from
scale and PDF uncertainties and also present differential cross
sections at NLO for $b {\overline b} h$ production in the case when
two $b$ quarks are tagged.

\section{Results for $t\bar{t}h$ Production}
\label{sec:djrw_tth}
\unboldmath 
The observation of a $t\bar{t}h$ final state will allow for the
measurement of the $t\bar{t}h$ Yukawa coupling.  If
$M_{h}\!\le\!130$~GeV, $pp\to t\bar{t}h$ is an important discovery
channel for a SM-like Higgs boson at the LHC
($\sqrt{s}\!=\!14$~TeV)~\cite{Beneke:2000hk,Drollinger:2001yy}.  Given
the statistics expected at the LHC, $pp\to t\bar{t}h$, with $h\to
b\bar{b},\tau^+\tau^-,W^+W^-,\gamma\gamma$ will be instrumental for
the determination of the couplings of the Higgs boson.  Precisions of
the order of 10-15\% on the measurement of the top quark Yukawa
coupling can be obtained with integrated luminosities of 100~fb$^{-1}$
per detector\cite{Zeppenfeld:2000td,Belyaev:2002ua,Maltoni:2002jr,Duhrssen:2004uu}.

\begin{figure}[t]
\begin{center}
\includegraphics[bb=150 500 430 700,scale=0.65]{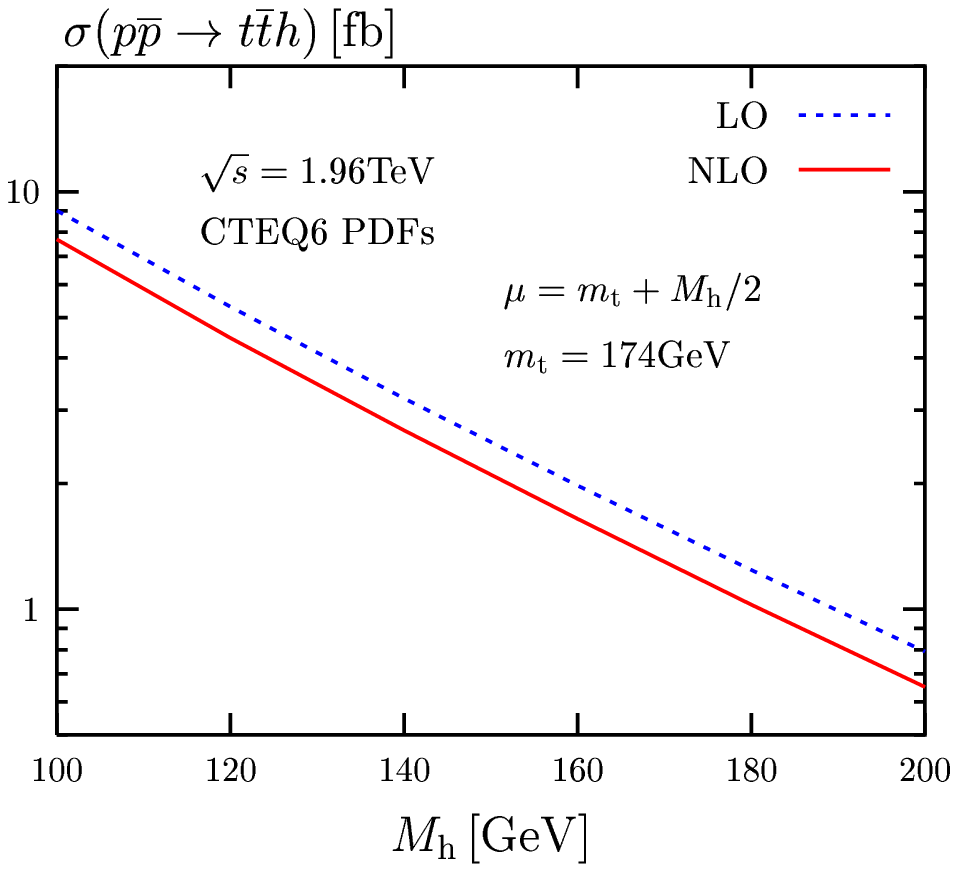} 
\includegraphics[bb=150 500 430 700,scale=0.65]{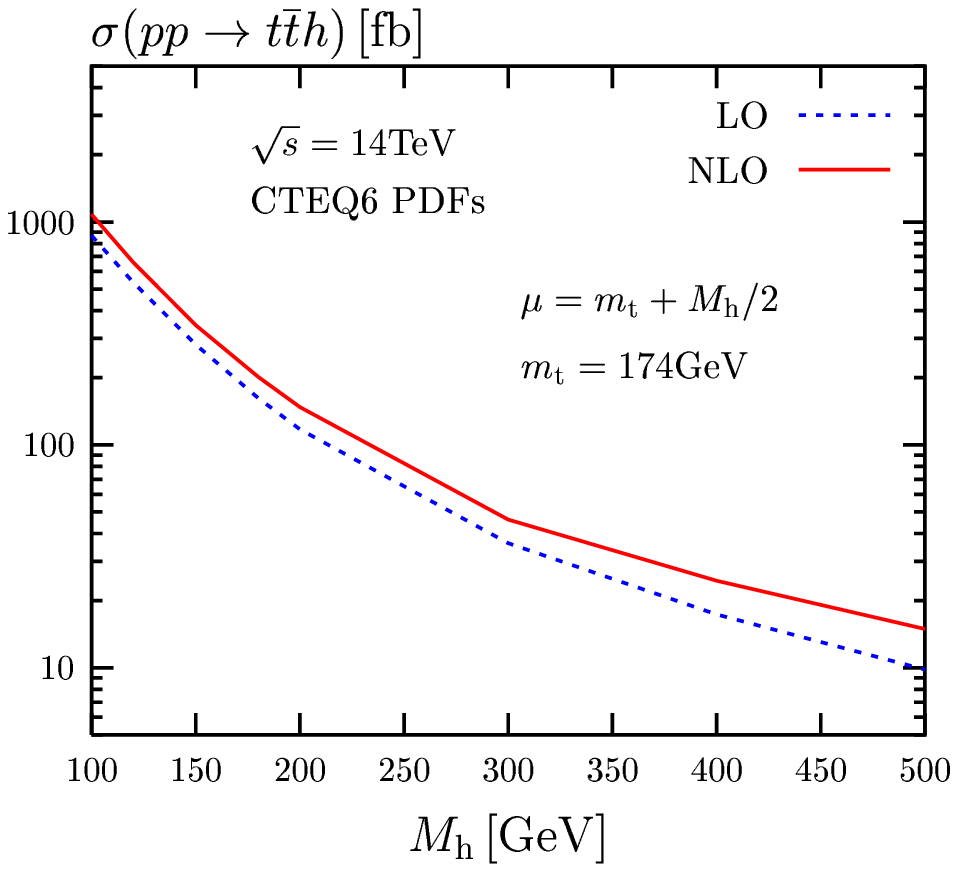} 
\vspace{0.5truecm}  
\caption[]{Total LO and NLO cross sections for $pp,p\overline{p}\to t\bar{t}h$
as functions of $M_h$, at $\sqrt{s} \!=\!1.96$~TeV and $\sqrt{s}
\!=\!14$~TeV, for $\mu\!=m_t+M_h/2$.}
\label{fg:tth_mhdep}
\end{center}
\end{figure}
\begin{figure}[t]
\begin{center}
\begin{tabular}{rl}
\includegraphics[scale=0.4]{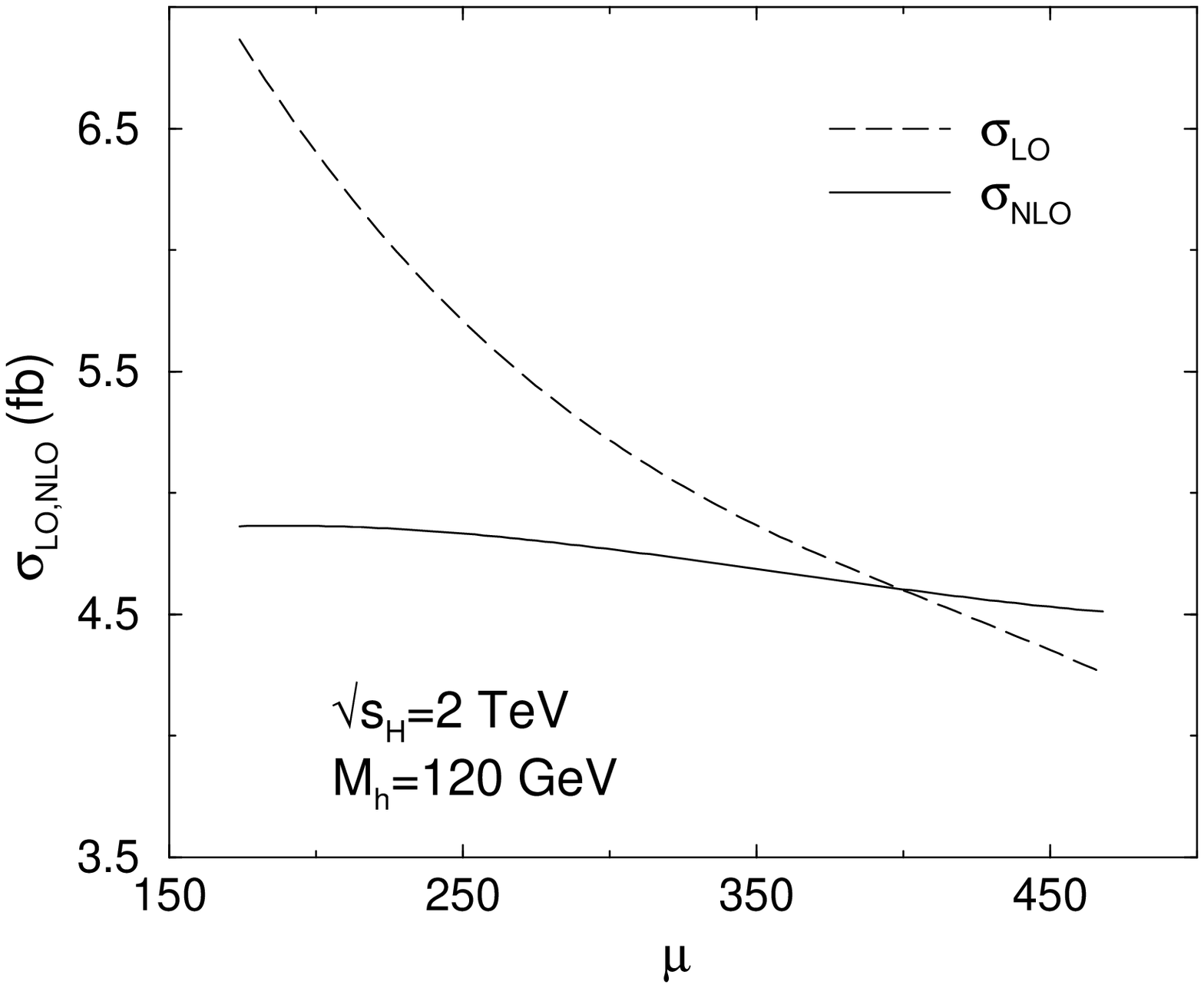} &
\includegraphics[scale=0.4]{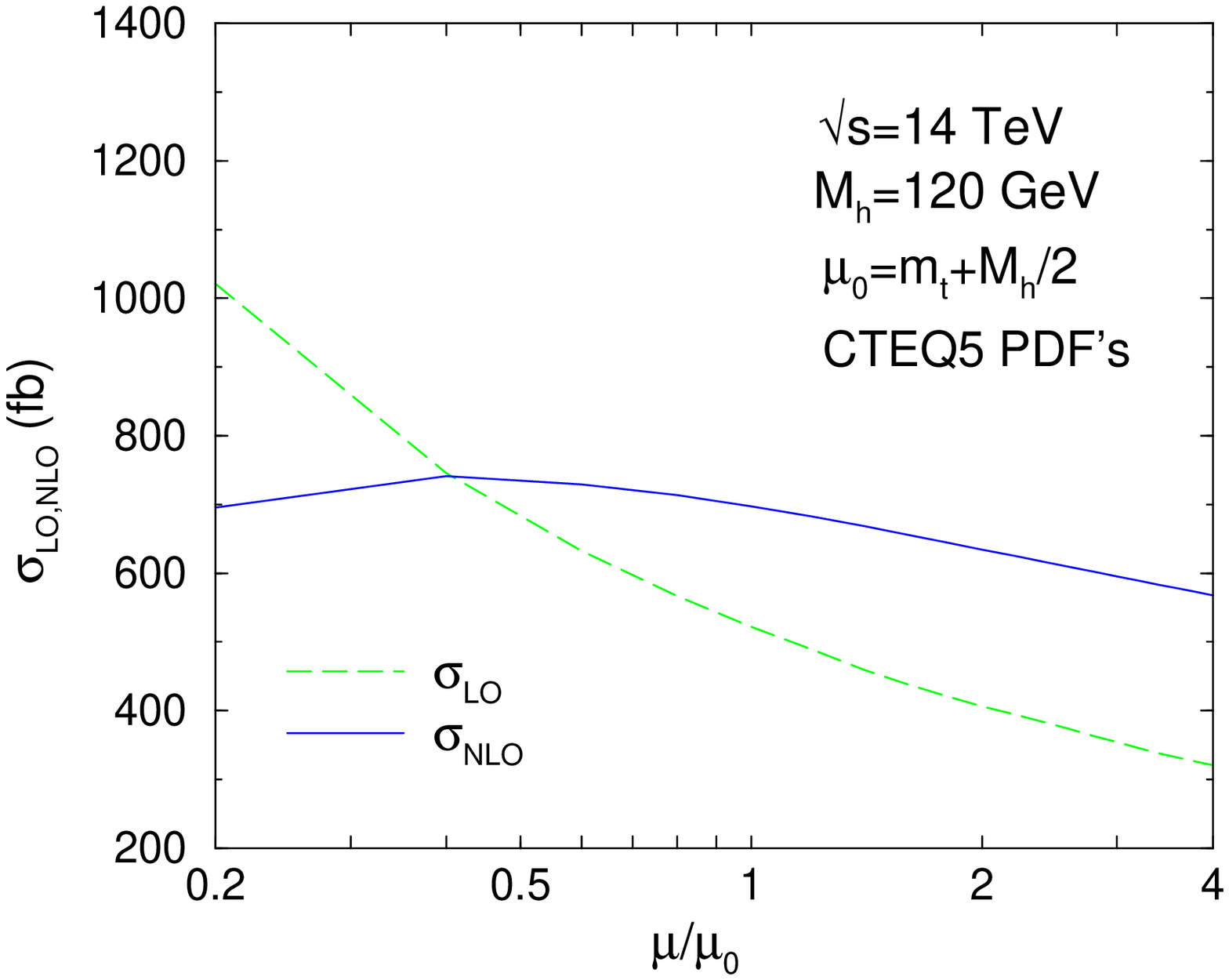}
\end{tabular} 
\caption[]{Dependence of $\sigma_{\sss LO,NLO}(pp,p\bar p\to t\bar{t}h)$ on 
  the renormalization/factorization scale $\mu$, at $\sqrt{s}\!=\!2$~TeV (l.h.s.)and
$\sqrt{s}\!=\!14$~TeV (r.h.s.), for $M_h\!=\!120$ GeV.}
\label{fg:tth_mudep}
\end{center}
\end{figure}

The impact of NLO QCD corrections on the total cross section for
$p\overline{p},pp\to t\bar{t}h$ production in the Standard Model is
illustrated in
Fig.~\ref{fg:tth_mhdep}\cite{Dawson:2003zu,Dawson:2002tg,Reina:2001sf,Reina:2001bc}
and Fig.~\ref{fg:tth_mudep}\cite{Dawson:2003zu,Reina:2001bc}. The
dependence of the total cross sections on the renormalization and
factorization scales is strongly reduced at NLO as shown in
Fig.~\ref{fg:tth_mudep}.  The numerical results at NLO are obtained
using CTEQ4M (Fig.~\ref{fg:tth_mudep} (l.h.s.)), CTEQ5M
(Fig.~\ref{fg:tth_mudep} (r.h.s.)), and CTEQ6M
(Fig.~\ref{fg:tth_mhdep}) parton distribution functions.  The NLO
cross section is evaluated using the 2-loop evolution of
$\alpha_s(\mu)$ with $\alpha_s^{NLO}(M_Z)=0.116$
(Fig.~\ref{fg:tth_mudep} (l.h.s.)) and $\alpha_s^{NLO}(M_Z)=0.118$
(Fig.~\ref{fg:tth_mudep} (r.h.s.)) and Fig.~\ref{fg:tth_mhdep}), and
$m_t=174$~GeV.  The renormalization/factorization scale dependence,
uncertainty on the PDFs, and the error on the top quark pole mass,
$m_t$, are estimated to give a 15-20$\%$ uncertainty.

\section{Results for $b\bar{b}h$ Production}
\label{sec:djrw_bbh}

The $b\bar bh$ production processes are only relevant discovery modes
in the MSSM with large $\tan\beta$. To a good approximation, the
predictions for the MSSM rates can easily be derived from the Standard
Model results by rescaling the Yukawa couplings\cite{Dawson:2005vi}.
The dominant MSSM radiative correction to $b\bar bh$ production can be
taken into account by including the MSSM corrections to the $b\bar bh$
vertex only, i.e. by replacing the tree level Yukawa couplings by the
radiative corrected ones. We follow the treatment of the program {\sc
FeynHiggs}~\cite{FeynHiggs:2005,Hahn:2004td} and take into account the
leading, $\tan\beta$ enhanced, radiative corrections that are
generated by gluino-sbottom and chargino-stop loops.  For large
$\tan\beta$, the bottom quark Yukawa coupling is enhanced and the top
quark Yukawa coupling coupling is strongly suppressed, resulting in a
MSSM $b\bar bh$ cross section that is about three orders of magnitude
larger than the Standard Model cross section.  For the Tevatron, we
calculate the production rates for the lightest MSSM Higgs boson,
$h^0$, while for the LHC we consider the rate for the heavier neutral
Higgs boson, $H^0$.\footnote{We assume $M_{SUSY}=1$~TeV,
$M_{\tilde{g}} =1$~TeV, $A_b=A_t=2$~TeV ($h^0$), $A_b=A_t=25$~GeV
($H^0$), $\mu=M_2=200$~GeV ($h^0$), and $\mu=M_2=1$~TeV ($H^0$).  For
$M_{h^0}=120$~GeV, the $bbh^0$ coupling is enhanced by a factor of
$33$ relative to the SM coupling, while for $M_{H^0}$ between $200$
and $800$~GeV, the $bbH^0$ coupling is enhanced by a factor of $27$
relative to the SM coupling.}

%
In the numerical evaluation of cross sections for the exclusive and
semi-inclusive channels ($b{\overline b}h$ and $bh+\bar{b}h$
production), it is required that the final state bottom quarks have
$p_T\!>\!20~$GeV and pseudorapidity $\mid\!\eta\!\mid<\!2.0$ for the
Tevatron and $\mid\!\eta\!\mid<\!2.5$ for the LHC. In the NLO real
gluon emission contributions, the final state gluon and bottom quarks
are considered as separate particles only if their separation in the
pseudorapidity-azimuthal angle plane, $\Delta
R\!=\!\sqrt{(\Delta\eta)^2+(\Delta\phi)^2}$, is larger than $0.4$. For
smaller values of $\Delta R$, the four momentum vectors of the two
particles are combined into an effective bottom/anti-bottom quark
momentum four-vector.

If not stated otherwise, the numerical results at NLO are obtained
using CTEQ6M PDFs, the 2-loop evolution of $\alpha_s(\mu)$ with
$\alpha_s^{NLO}(M_Z)=0.118$, and the $\overline{MS}$ renormalization
scheme for the bottom quark mass and Yukawa coupling with 2-loop
renormalization group improved $\overline{MS}$ masses. The bottom quark
pole mass is chosen to be $m_b=4.62$~GeV. 

\subsection{Total Cross Sections for  $b\bar bh$ Production}
\label{subsec:djrw_0btag}
\begin{figure}[btp]
\begin{center}
\begin{tabular}{rl}
\hskip 2.2in
\includegraphics[bb=90 700 524 13,scale=0.25,angle=-90]{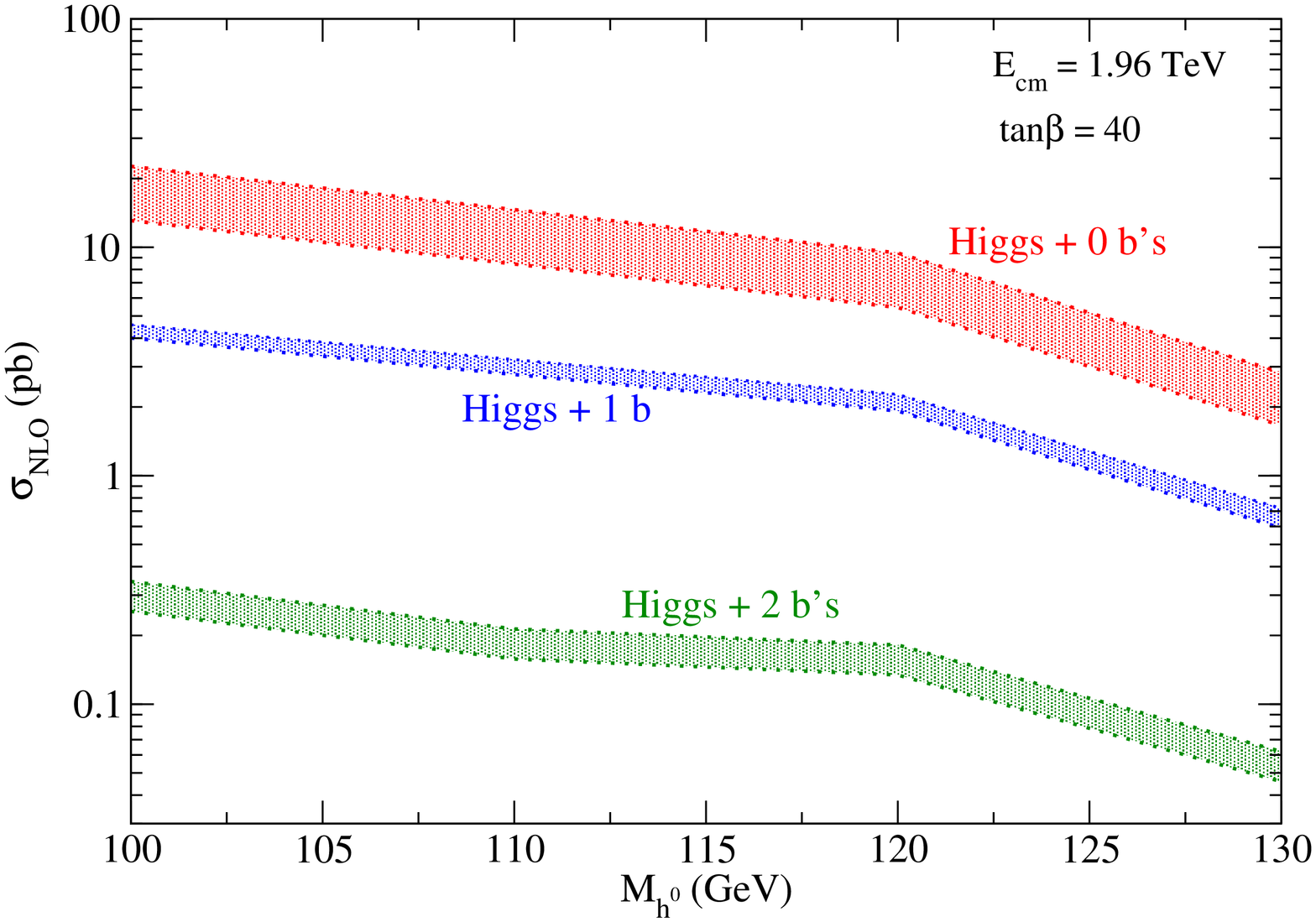} & 
\hskip 3in
\includegraphics[bb=90 700 524 12 ,scale=0.25,angle=-90]{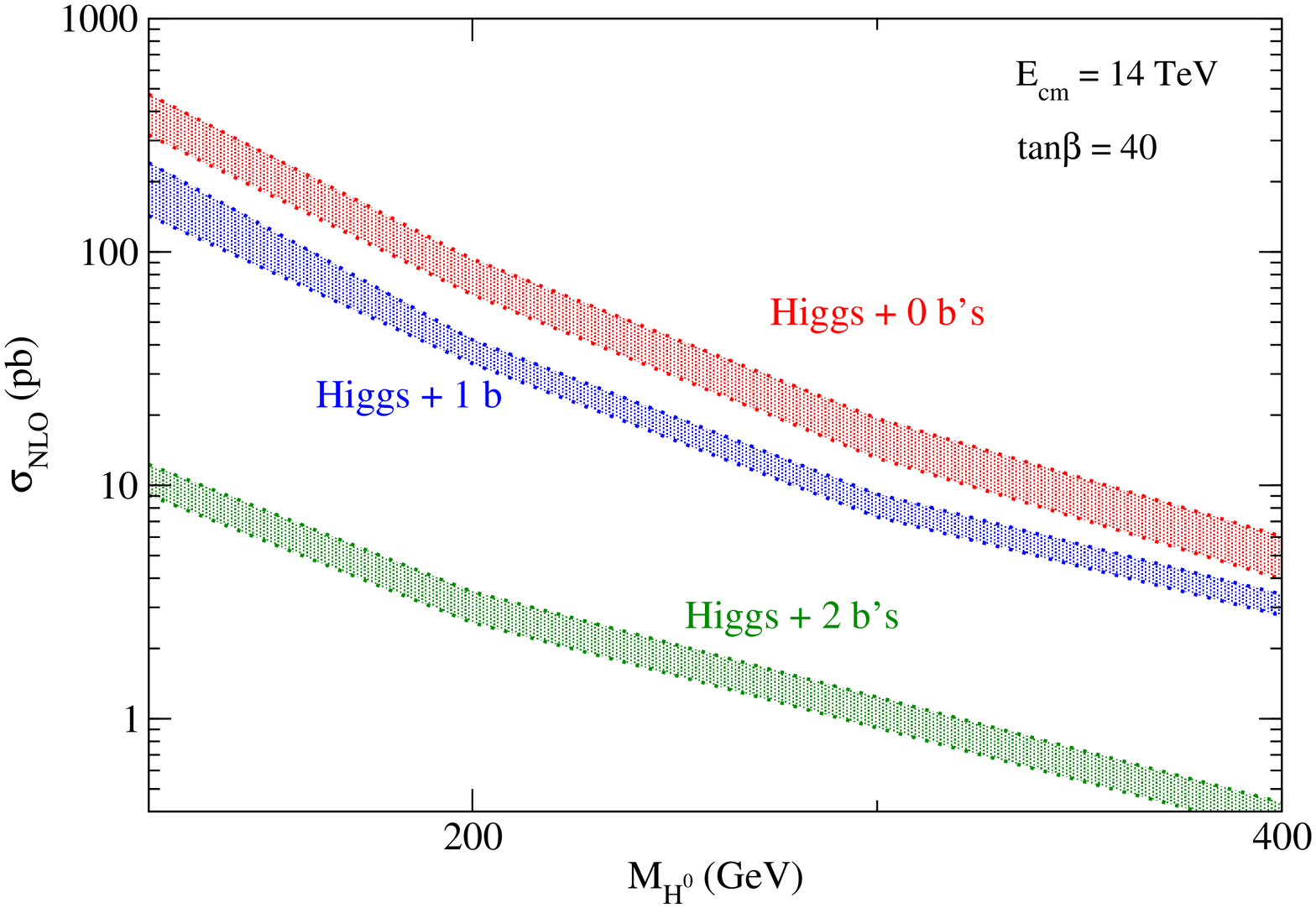}
\end{tabular}
\vspace*{8pt}
\caption[]{Total cross sections for $pp,p\bar p\rightarrow b\bar{b}h$
  in the MSSM in the 4FNS at NLO for the Tevatron and the LHC in 
the MSSM with $\tan\beta=40$ and with 0,1 or 2 $b$
quarks identified.  The Tevatron (LHC)  plot is for the 
lightest (heaviest) neutral Higgs boson, $h^0$ ($H^0$).
  The error bands have been obtained by varying
  the renormalization and factorization scales as
  described in the text.}
\label{fg:tot}
\end{center}
\end{figure}

We present total cross section results at NLO in the 4FNS in
Fig.~\ref{fg:tot} for associated $b\bar b$ Higgs production in the
MSSM with $\tan\beta=40$.  The bands represent the theoretical
uncertainty due to the residual scale dependence. They have been
obtained by varying the renormalization ($\mu_r$) and factorization
($\mu_f$) scales independently from $\mu_0/4$ to $\mu_0$, where
$\mu_0\!=\!m_b+M_h/2$.

If the outgoing bottom quarks cannot be observed then the dominant
MSSM Higgs production process at large $\tan\beta$ is $gg\rightarrow
(b\bar{b})h$ (the curve labelled '0 b').  The inclusive cross section
is experimentally relevant only if the Higgs boson can be detected
above the background without tagging bottom quarks.  At the LHC, this
process can be identified at large $\tan\beta$ by the decays to
$\mu^+\mu^-$ and $\tau^+\tau^-$ for the heavy Higgs bosons, $H^0$ and
$A^0$, of the MSSM.  At the Tevatron this process, with
$h^0\rightarrow \tau^+\tau^-$, has been used to search for the neutral
MSSM Higgs boson.
If a single bottom quark is tagged then the final state is $bh$ or
$\bar{b}h$ (the curve labelled '1 b'). Although requiring a $b$ quark in
the final state significantly reduces the rate, it also reduces the
background.  A recent Tevatron study~\cite{Abazov:2005yr} used the search
for neutral MSSM Higgs bosons in events with three bottom quarks in
the final state ($bh^0+\bar{b}h^0$ production with $h^0\rightarrow
b\bar{b}$) to impose limits on the $\tan\beta$ and $M_{A^0}$ parameter
space.
 
Finally, we show the fully exclusive cross sections for $b\bar{b}h$
production, where both the outgoing $b$ and $\bar{b}$ quarks are
identified (the curve labelled '2 b').  The exclusive measurement
corresponds to the smallest cross section, but it also has a
significantly reduced background. Moreover, both the exclusive and
semi-inclusive $b\bar bh$ production modes are the only ones that can
unambiguously measure the bottom quark Yukawa coupling.

\subsection{Differential Cross Sections for $b\bar bh $ Production}

\begin{figure}[t]
\begin{center}
\includegraphics[bb=150 500 430 700,scale=0.5]{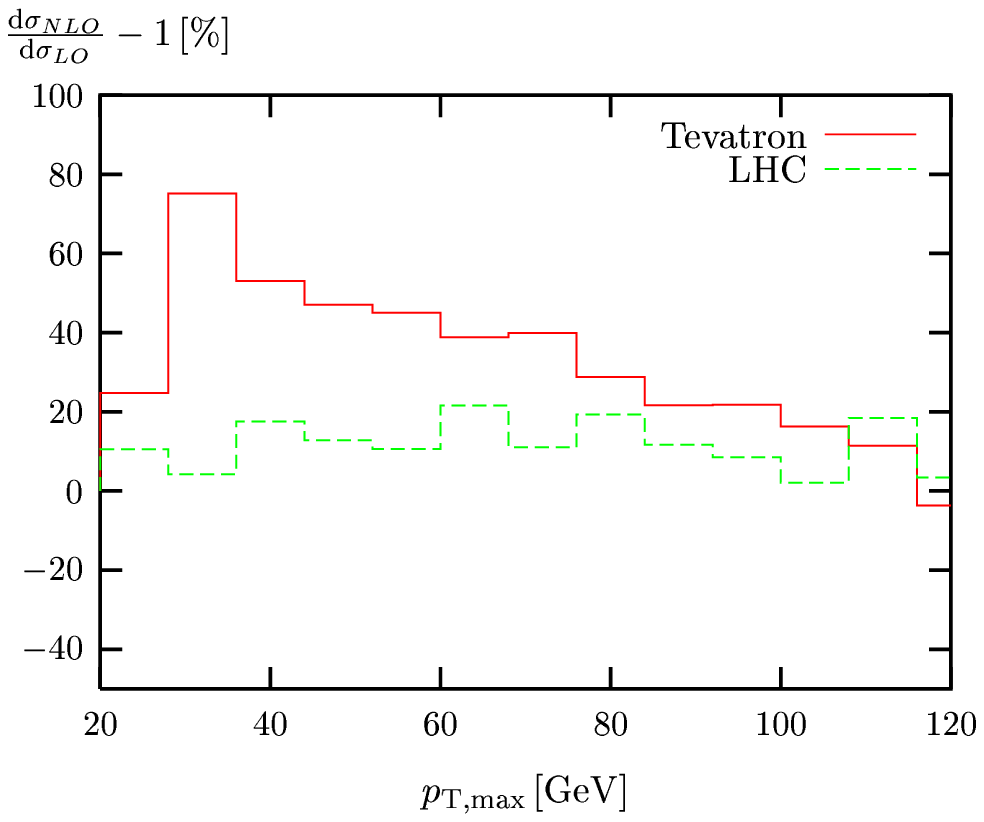} 
\includegraphics[bb=150 500 430 700,scale=0.5]{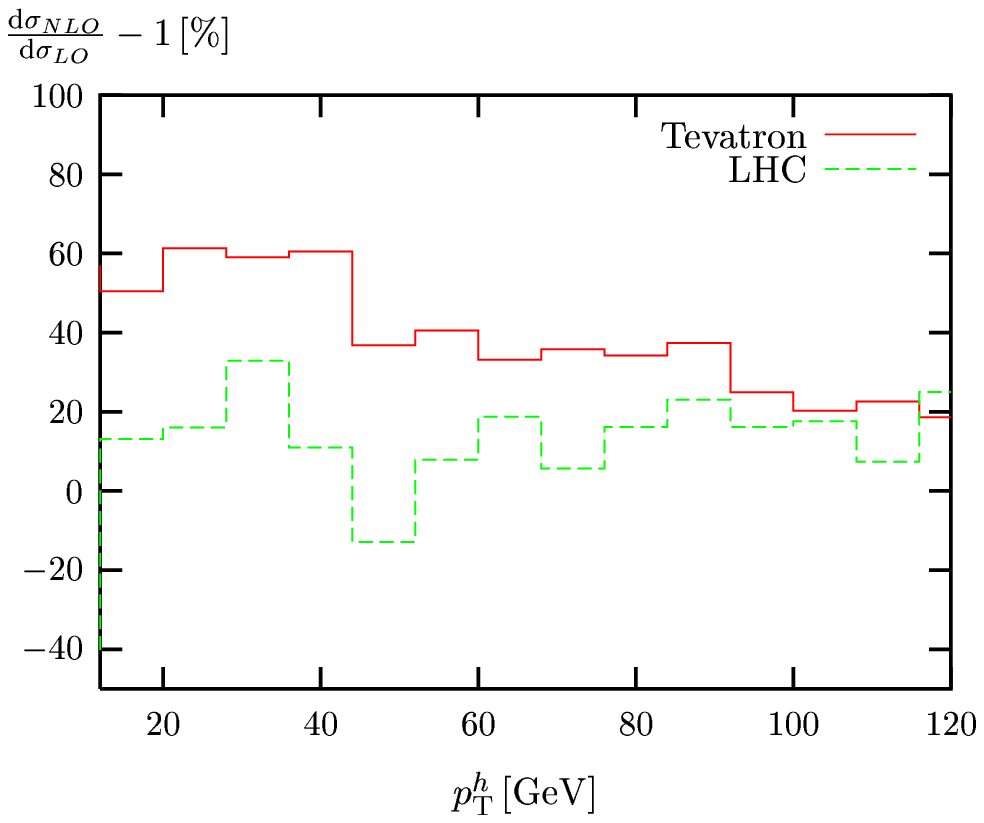} 
\caption[]{The relative corrections $d\sigma_{\sss NLO}/d \sigma_{\sss
LO}-1$ for the $p_T$ distribution of the bottom or anti-bottom quark
with the largest $p_T$ ($p_{T,max}$) (left) and of the SM Higgs boson
($p_{T}^h$) (right) to $b\bar b h$ production in the SM at the
Tevatron (with $\sqrt{s}\!=\!2$~TeV and $\mu\!=\!2 \mu_0$) and the LHC
(with $\sqrt{s}\!=\!14$~TeV and $\mu\!=\!4 \mu_0$).}
\label{fg:bbh_ptrel}
\end{center}
\end{figure}

\begin{figure}[t]
\begin{center}
\includegraphics[bb=150 500 430 700,scale=0.5]{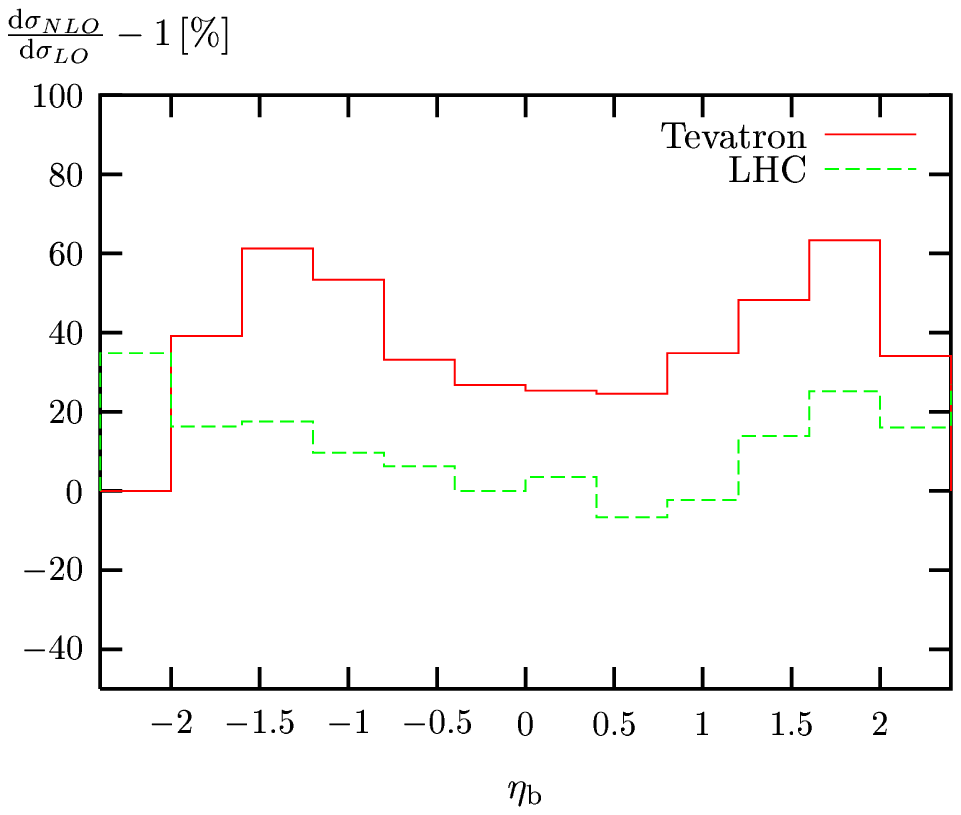} 
\includegraphics[bb=150 500 430 700,scale=0.5]{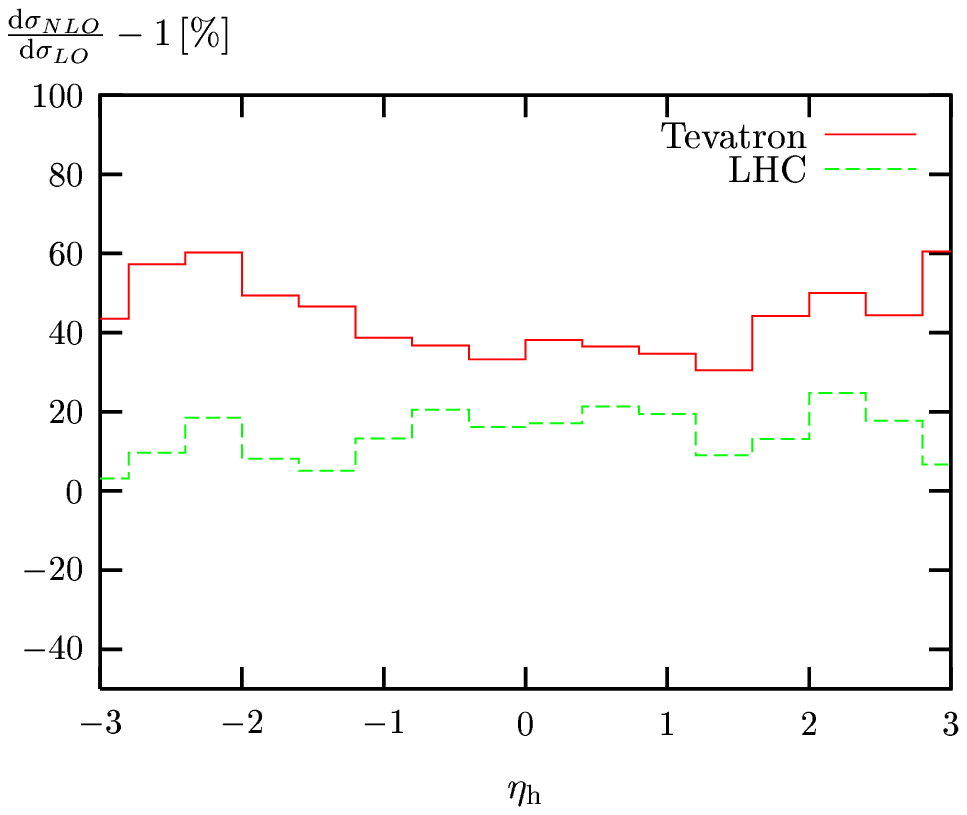} 
\caption[]{The relative corrections $d\sigma_{\sss NLO}/d \sigma_{\sss
LO}-1$ for the $\eta$ distribution of the bottom quark $\eta_b$ (left)
and of the SM Higgs boson ($\eta_{h}$) (right) to $b\bar b h$
production in the SM at the Tevatron (with $\sqrt{s}\!=\!2$~TeV and
$\mu\!=\!2 \mu_0$) and the LHC (with $\sqrt{s}\!=\!14$~TeV and
$\mu\!=\!4 \mu_0$).}
\label{fg:bbh_etarel}
\end{center}
\end{figure}

In assessing the impact of the NLO corrections it is particularly
interesting to study the kinematic distributions.  In
Figs.~\ref{fg:bbh_ptrel} and \ref{fg:bbh_etarel} we illustrate the
impact of NLO QCD corrections on the transverse momentum and
pseudorapidity distribution of the SM Higgs boson and the bottom quark
by showing the relative correction, $d\sigma_{NLO}/d\sigma_{LO}-1$ (in
percent) for the exclusive case ($b {\overline b} h$ where both $b$
quarks are observed).  For the renormalization/factorization scale we
choose $\mu=2 \mu_0$ at the Tevatron and $\mu=4 \mu_0$ at the LHC, with
$\mu_0=m_b+M_h/2$, and use the CTEQ5 set of PDFs.  As
can be seen, the NLO QCD corrections can considerably affect the shape
of kinematic distributions, and their effect cannot be obtained from
simply rescaling the LO distributions with a K-factor of $\sigma_{\sss
NLO}/\sigma_{\sss LO}\!=\!1.38\pm 0.02$ (Tevatron, $\mu\!=\!2\mu_0$)
and $\sigma_{\sss NLO}/\sigma_{\sss LO}\!=\!1.11\pm 0.03$ (LHC,
$\mu\!=\!4\mu_0$).\footnote{The kinematic distributions have been
calculated within the Standard Model and using the on-shell scheme for
the definition of the $b$ quark mass, but we see a similar behavior
when using the $\overline{MS}$ bottom quark Yukawa coupling.}

\subsection{PDF and Renormalization/Factorization Scale Uncertainties}
\label{sec:djrw_pdfs}
A major source of theoretical uncertainty for cross section
predictions comes from the PDFs.  We study the uncertainties of
semi-inclusive $bh$ production rates from the uncertainties in
the PDFs using the CTEQ PDF sets\cite{Pumplin:2002vw}.  First, the
central value cross section $\sigma_0$ is calculated using the global
minimum PDF (i.e. CTEQ6M).  The calculation of the cross section is
then performed with the additional 40 sets of PDFs to produce 40
different predictions, $\sigma_i$.  For each of these, the deviation
from the central value is calculated to be $\Delta\sigma_i^{\pm} =
|\sigma_i-\sigma_0|$ when ${\sigma_i}_{<}^{>} \sigma_0$.  Finally, to
obtain the uncertainties due to the PDFs the deviations are summed
quadratically as $\Delta\sigma^{\pm} = \sqrt{ \sum_i
{\Delta\sigma^{\pm}_{i}}^{2}}$ and the cross section including the
theoretical uncertainties arising from the PDFs is quoted as
$\sigma_0|^{+\Delta\sigma^{+}}_{-\Delta\sigma^{-}}$.

\begin{figure}[t]
\begin{center}
\includegraphics[scale=0.22,angle=-90]{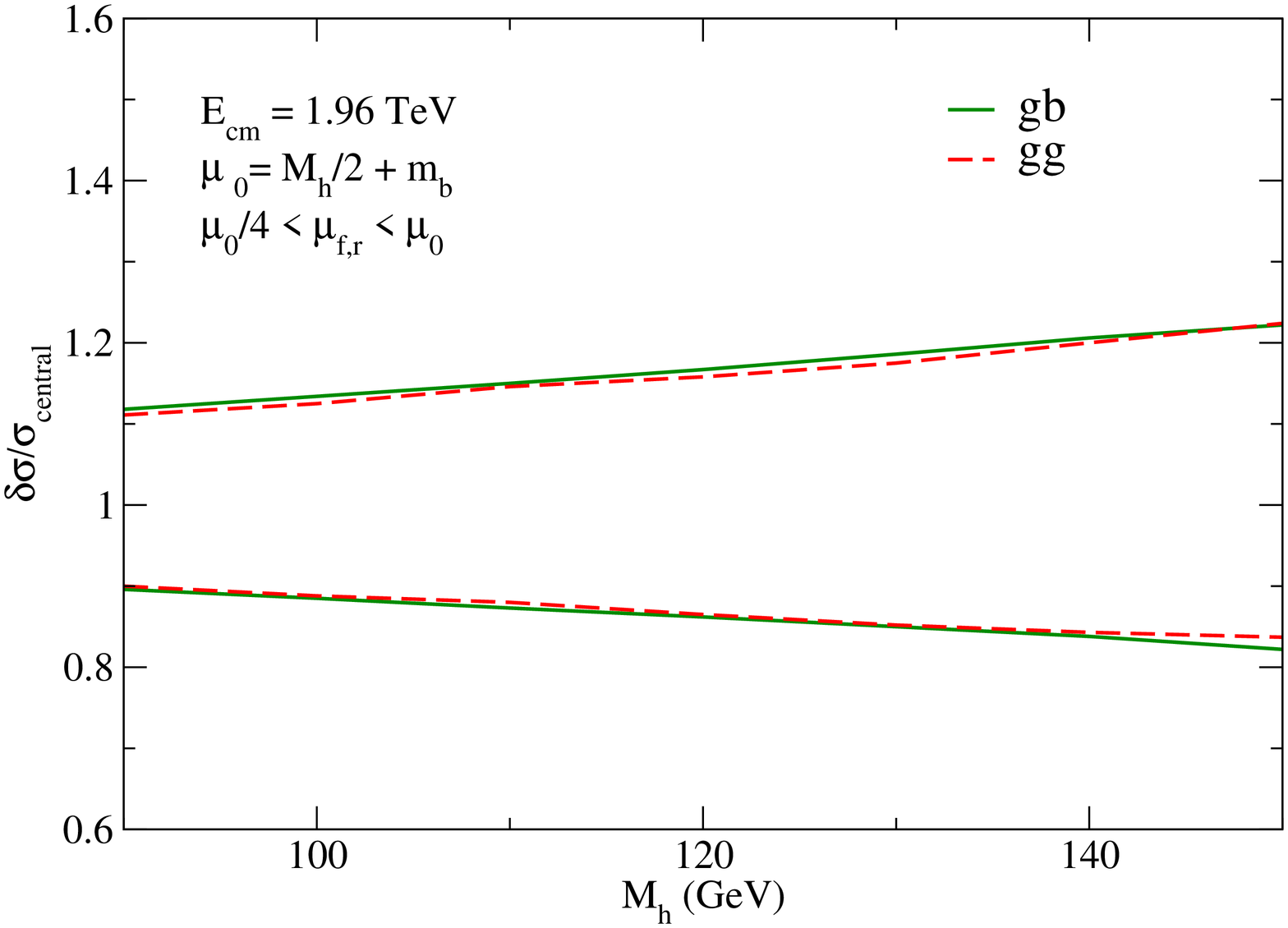}
\includegraphics[scale=0.22,angle=-90]{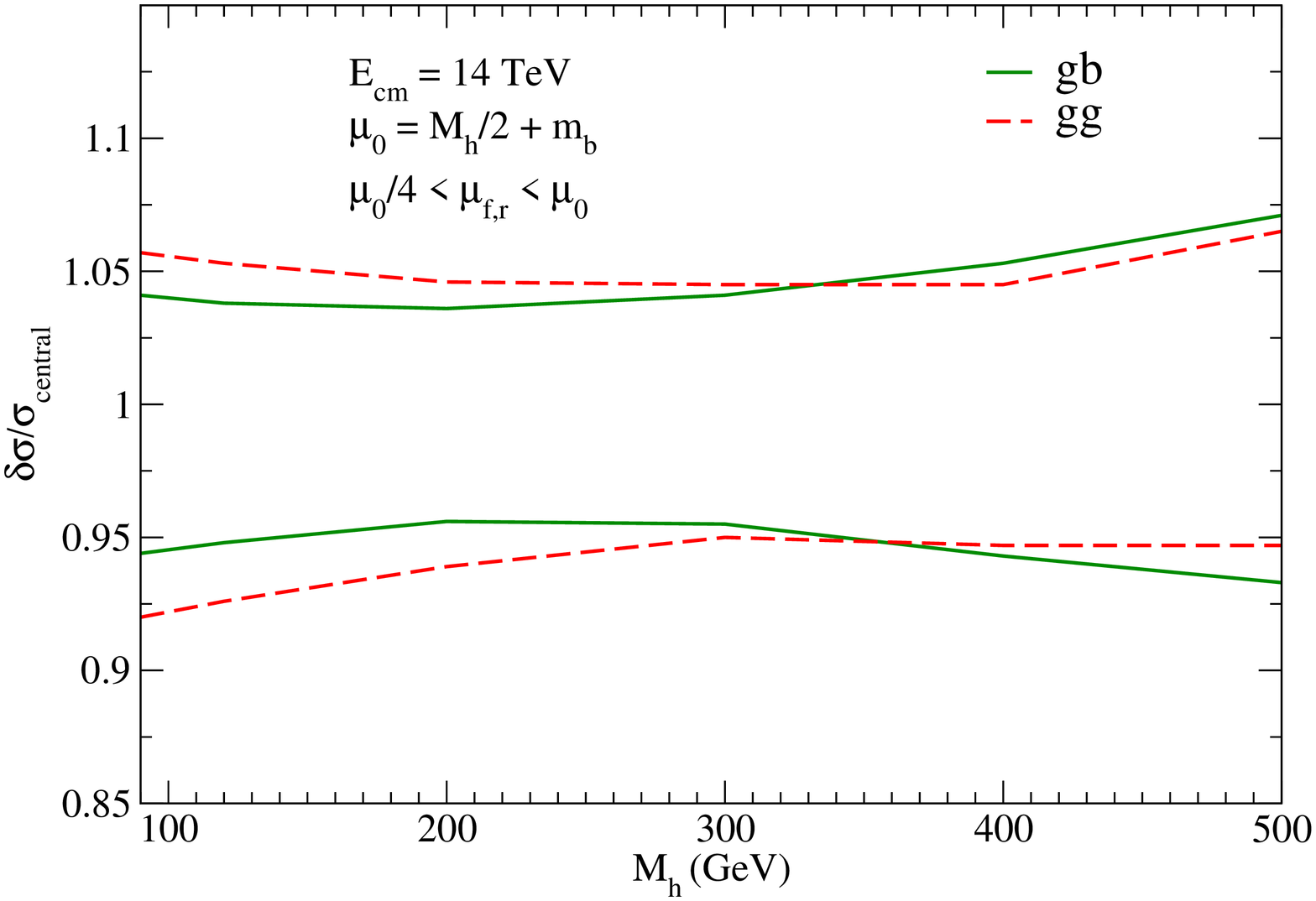} 
\caption[]{Normalized cross sections for Higgs production with one $b$
jet at the Tevatron (l.h.s) and the LHC (r.h.s) showing the uncertainty
from PDFs for both the $gg$ (4FNS) and $bg$ (5FNS) initial states.}
\label{fg:ggvbg}
\end{center}
\end{figure}

In Fig.~\ref{fg:ggvbg}, we plot the normalized total SM NLO cross
sections for semi-inclusive $bh$ production, calculated in the
5FNS ($bg\rightarrow bh$) as implemented in MCFM~\cite{MCFM:2004} and
in the 4FNS ($gg\rightarrow b(\bar{b})h$), and compare their
respective uncertainties due to the PDFs.  We see that, at both the
Tevatron and the LHC, the PDF uncertainties are almost identical for
both the $gg$ and $bg$ initial states.

\begin{figure}[t]
\begin{center}
\begin{tabular}{rl}
\includegraphics[scale=0.25,angle=-90]{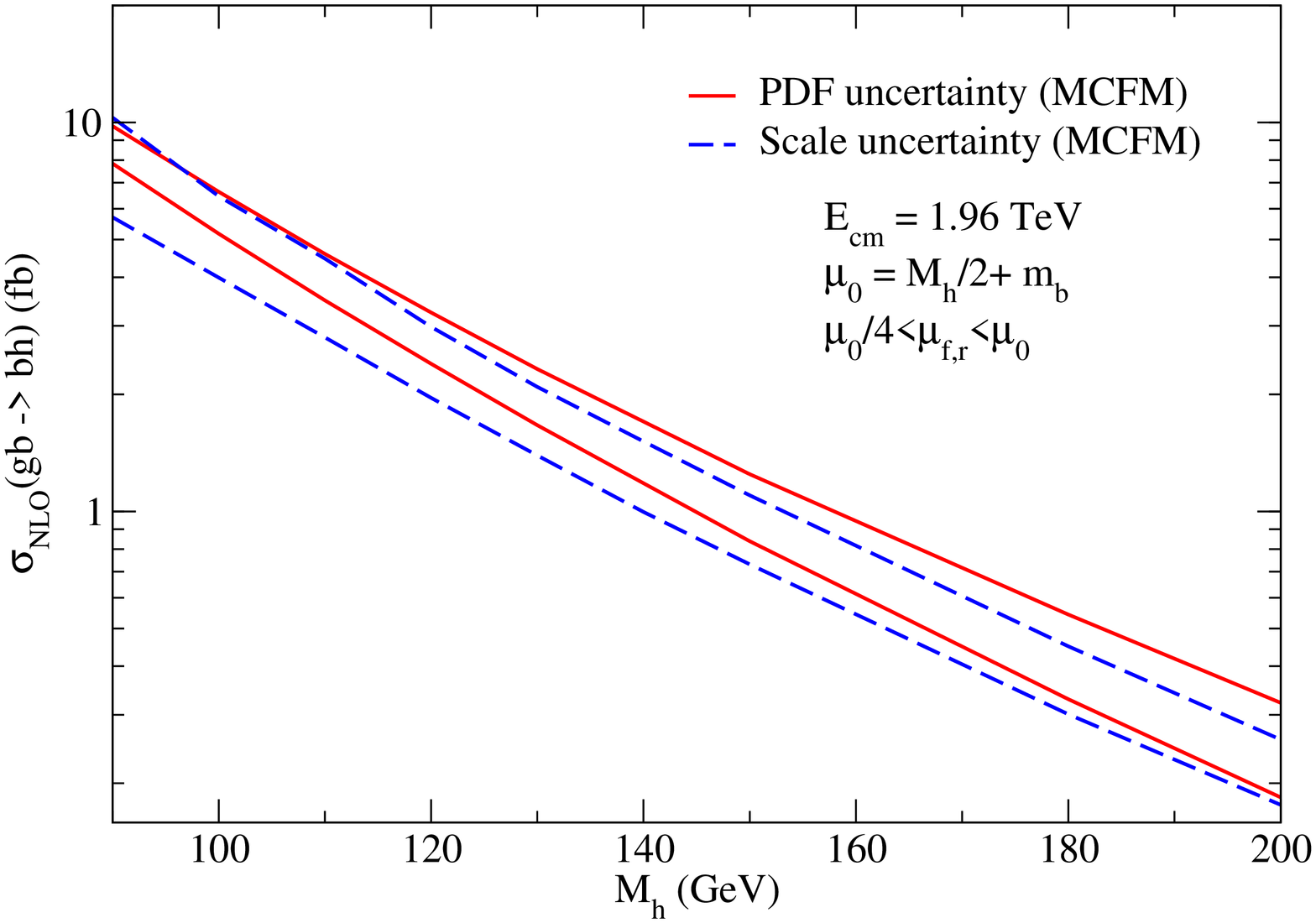} &
\includegraphics[scale=0.25,angle=-90]{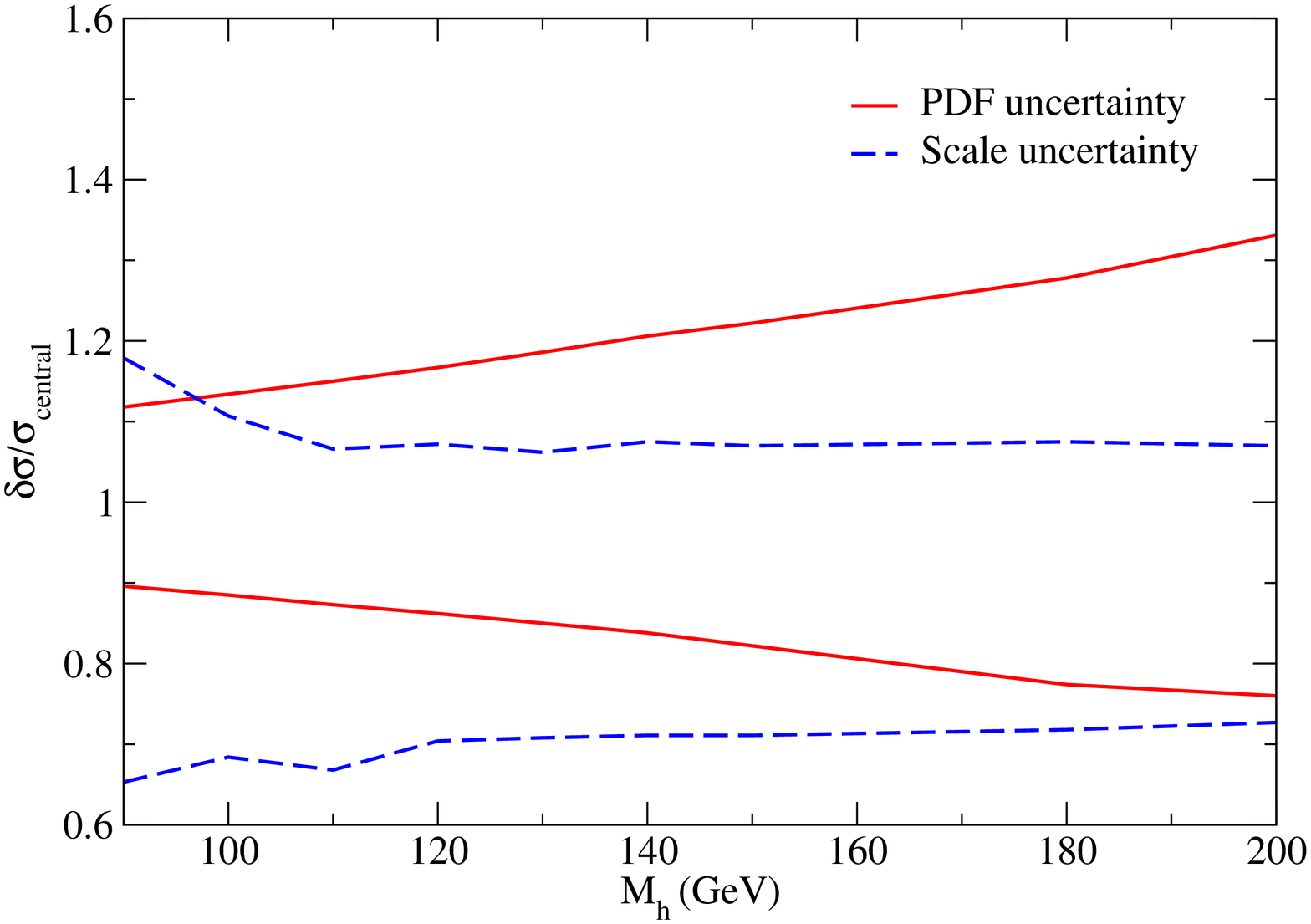}
\end{tabular} 
\caption[]{Comparison between theoretical uncertainties due to scale
dependence and uncertainties arising from the PDFs at the Tevatron
for
semi-inclusive $bh$ production in the Standard Model.
In the right-hand plot, both uncertainty bands have been normalized to the 
central value of the total cross section $\sigma_0$.}
\label{fg:tevPDF}
\end{center}
\end{figure}

\begin{figure}[t]
\begin{center}
\begin{tabular}{rl}
\includegraphics[scale=0.25,angle=-90]{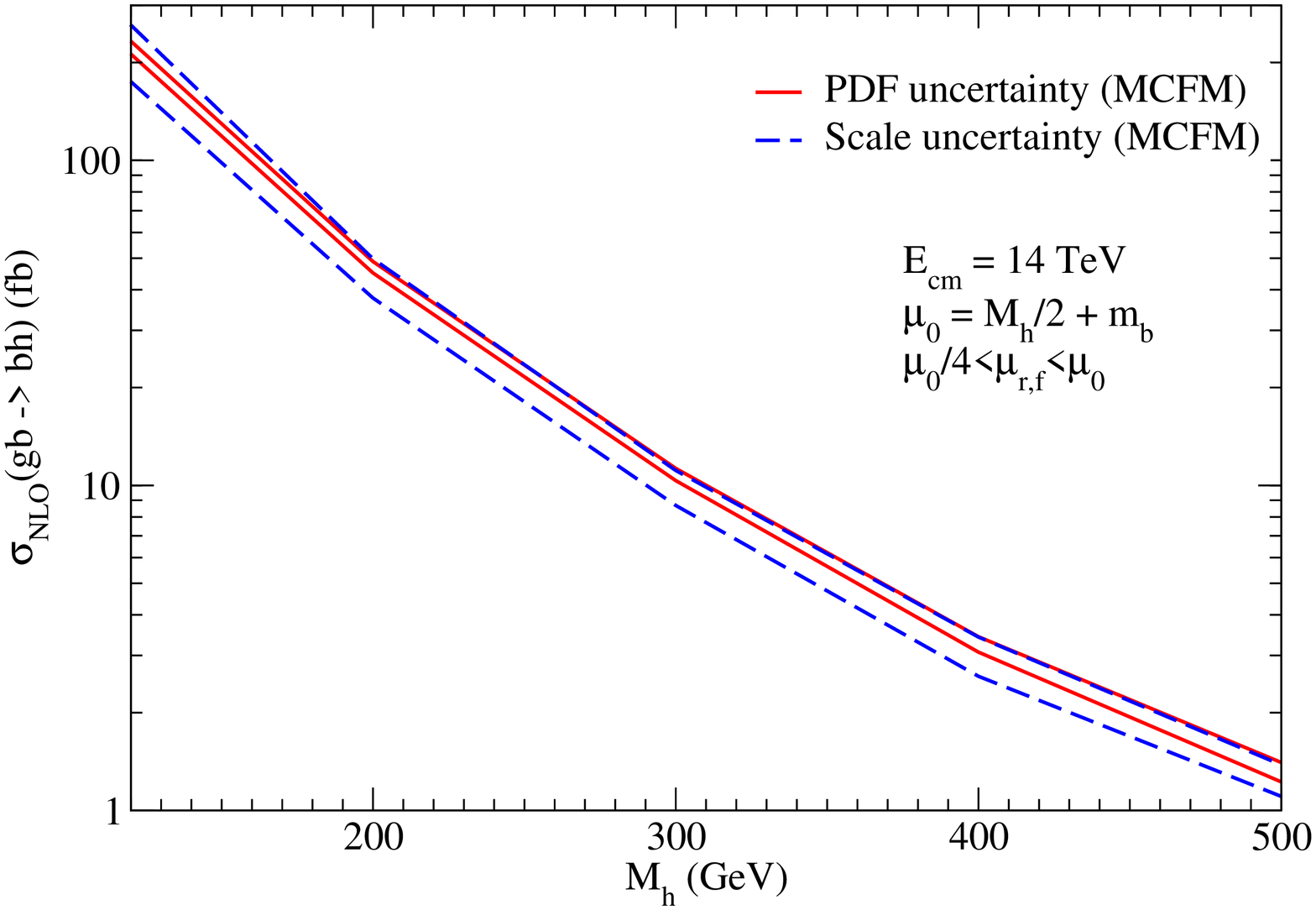} &
\includegraphics[scale=0.25,angle=-90]{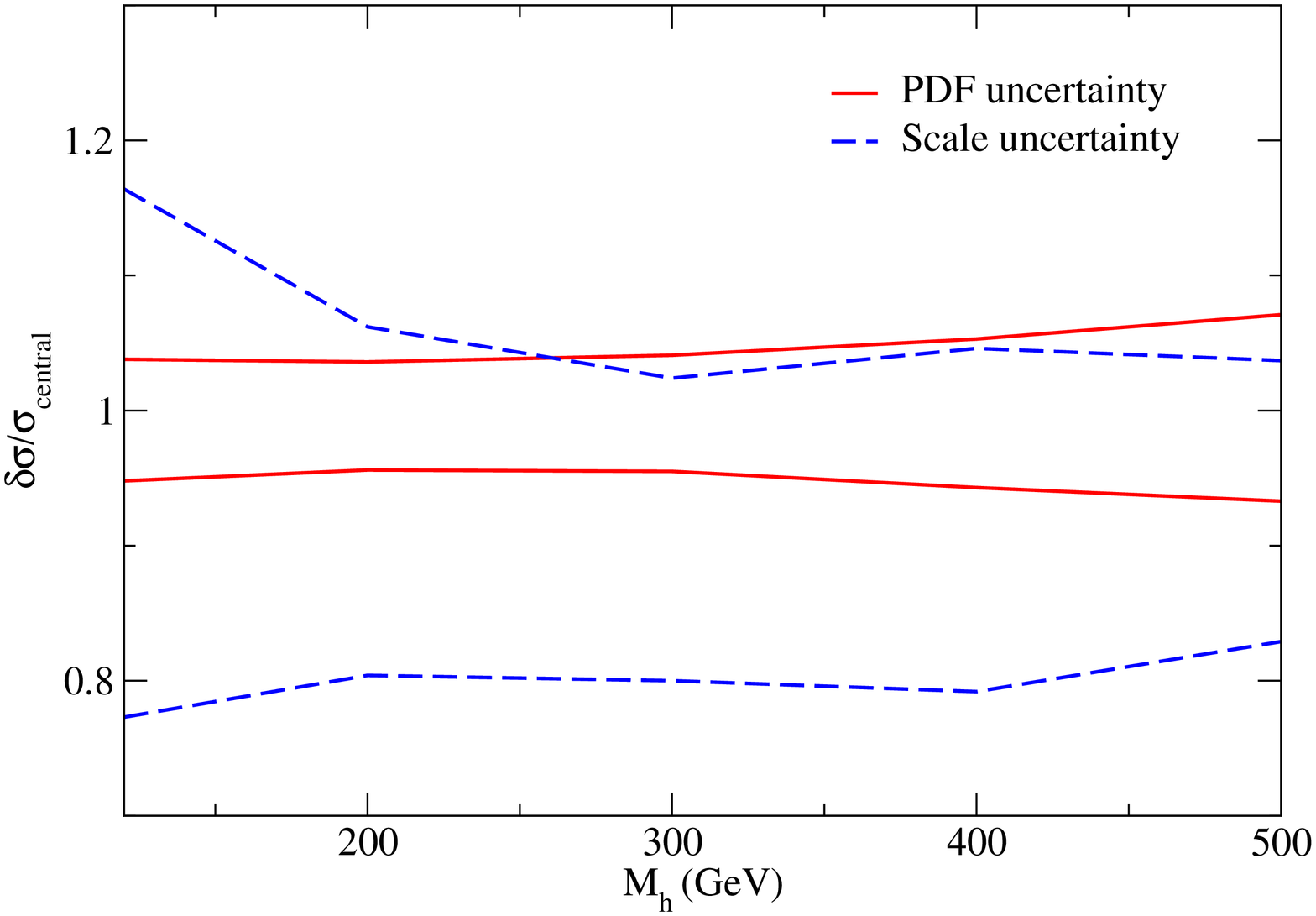}
\end{tabular} 
\caption[]{Comparison between theoretical uncertainties due to scale
dependence and uncertainties arising from the PDFs at the LHC for
semi-inclusive $bh$ production in the Standard Model.
In the bottom plot, both uncertainty bands have been normalized to the 
central value of the total cross section $\sigma_0$.}
\label{fg:lhcPDF}
\end{center}
\end{figure}
In Figs.~\ref{fg:tevPDF} and~\ref{fg:lhcPDF} we 
compare the uncertainties from residual scale dependence and the PDFs
on the example of $bg\rightarrow bh$ (5FNS)   
at the Tevatron and LHC
respectively\cite{Dawson:2005vi}.  Here, we perform the comparison 
for both the total cross
section (left) and the total cross section normalized to the central
value calculated with CTEQ6M (right). Similar results are obtained in the 4FNS.

From Fig.~\ref{fg:lhcPDF} one can see that, at the LHC, the
theoretical uncertainty is dominated by the residual scale dependence.
Due to the large center of mass (c.o.m.) energy of the LHC, the gluons
and bottom quarks in the initial state have small momentum fraction
($x$) values and, hence, small PDF uncertainties typically in the
5-10\% range.

In contrast, due to the smaller c.o.m. energy, the PDF uncertainties
at the Tevatron (Fig.~\ref{fg:tevPDF}) are comparable and even larger
than the uncertainties due to residual scale dependence over the full
Higgs mass range.  The smaller c.o.m. energy results in higher-$x$
gluons and bottom quarks in the initial state which corresponds to
large PDF uncertainties in the 10-30\% range.

\section{Conclusion}

The NLO cross sections for $t{\overline t}h$ and $b{\overline b}h$
have been presented for the Tevatron and the LHC with emphasis on the
renormalization/factorization scale and PDF dependences.
\section*{Acknowledgments}
The work of S.D. and C.J. ( L.R) is supported in part by the U.S.
Department of Energy under grant DE-AC02-98CH10886
(DE-FG-02-91ER40685). The work of D.W. is supported in part by the
National Science Foundation under grant No.~PHY-0244875.
%
\bibliographystyle{qqh}
\bibliography{qqh}

\end{document}